\documentclass[runningheads]{llncs}
\usepackage[english]{babel}
\usepackage{float}

\usepackage{array}
\usepackage{amsmath}
\usepackage{amssymb}
\usepackage{xspace}
\usepackage{mathpartir}

\usepackage{caption}
\usepackage{booktabs}
\usepackage{multirow}

\usepackage[frozencache]{minted}
\usepackage{graphicx}
\usepackage[export]{adjustbox}
\usepackage{pgfplotstable}
\usepackage{pgfplots}
\pgfplotsset{width=9cm,compat=1.8}
\graphicspath{{./graphics/}}

\newcommand{\semeq}[0]{\triangleq}

\newcommand{\stmt}[0]{s}
\newcommand{\stmts}[0]{\mathbb{S}}
\newcommand{\exprs}[0]{\mathbb{E}}
\newcommand{\vals}[0]{\mathbb{V}}
\newcommand{\vars}[0]{\mathbb{X}}
\newcommand{\ints}[0]{\mathbb{I}}

\newcommand{\compfun}[0]{\mathcal{C}}

\newcommand{\ecompiler}[1]{\mathcal{C}_e(#1)}
\newcommand{\compiler}[1]{ \compfun \left(#1\right)}
\newcommand{\getId}[0]{\mathcal{I}d}
\newcommand{\idx}[0]{idx}
\newcommand{\freshV}[0]{{\sf fresh}}

\newcommand{\str}[0]{str}
\newcommand{\auxV}[0]{aux}

\newcommand{\isBool}[0]{{\sf isBool}}
\newcommand{\ifstmt}[3]{\mintinline{c}{if}\,(#1)\,\{#2\}\,\mintinline{c}{else}\,\{#3\}}
\newcommand{\whilestmt}[2]{\mintinline{c}{while}\,(#1)\,\{#2\}}

\let\smallish\small

\newcommand{\hbra}{
\hbox to \linewidth{\vrule width0.3mm height 1.8mm depth-0.3mm
    \leaders\hrule height1.8mm depth-1.5mm\hfill
    \vrule width0.3mm height 1.8mm depth-0.3mm}}
\newcommand{\hket}{
\hbox to \linewidth{\vrule width0.3mm height1.5mm
    \leaders\hrule height0.3mm\hfill
    \vrule width0.3mm height1.5mm}}
\newcommand{\ratio}{.375}

\usepackage[linesnumbered, ruled, vlined]{algorithm2e}
\DontPrintSemicolon \SetKwData{false}{false}
\SetKwData{true}{true}
\SetKwData{and}{and}
\SetKwInOut{Input}{input}
\SetKwInOut{Output}{output}

\DeclareMathOperator*{\summaries}{\mathcal{S}}
\DeclareMathOperator*{\reals}{\mathbb{R}}

\DeclareMathOperator*{\hittingsets}{\textsf{HS}}

\DeclareMathOperator*{\ntests}{\text{n}}

\newcommand{\measure}{M}

\newcommand{\getCore}{\textsf{extractCore}}

\newcommand{\ABSTRACTION}{\Gamma}
\newcommand{\maxheuristic}{HeuristicMax}

\newcommand{\order}{IncOrder}

\newcommand{\testselector}{\textsc{TestSelector}\xspace}
\newcommand{\maxtests}{\textsc{MaxTests}\xspace}
\newcommand{\seesaw}{\textsc{Seesaw}\xspace}

\begin{document}
\title{\testselector: Automatic Test Suite Selection for Student Projects --- Extended Version}
\titlerunning{TestSelector}
\author{
  Filipe Marques\inst{1,2} \and Ant\'onio Morgado\inst{1} \and \\ Jos\'e Fragoso Santos\inst{1,2} \and Mikol\'a\v s Janota\inst{3} }
\authorrunning{F.\,Marques et al.}
\institute{
  INESC-ID, Lisbon, Portugal \\
\and Instituto Superior T\'ecnico, University of Lisbon, Portugal \\
\and
  Czech Technical University in Prague, Czechia }
\maketitle              \begin{abstract}
Computer Science course instructors routinely have to create comprehensive test suites to 
assess programming assignments. The creation of such test suites is typically not trivial as 
it involves selecting a limited number of tests from a set of (semi-)randomly generated ones. 
Manual strategies for test selection do not scale when considering large testing inputs needed, for 
instance, for the assessment of algorithms exercises. To facilitate this process, we present \testselector, 
a new framework for automatic selection of optimal test suites for student projects. The key 
advantage of \testselector over existing approaches is that it is easily extensible with arbitrarily 
complex code coverage measures, not requiring these measures to be encoded into the logic of an 
exact constraint solver. We demonstrate the flexibility of \testselector by extending it with support for a 
range of classical code coverage measures and using it to select test suites for a number of real-world 
algorithms projects, further showing that the selected test suites outperform randomly selected ones in 
finding bugs in students' code. 

\end{abstract}

\section{Introduction}\label{sec:intro}

Computer science course instructors routinely have to create comprehensive test suites to automatically assess programming assignments. 
It not uncommon for these test suites to have to be created before students actually submit their solutions. 
This is, for instance, the case when students are allowed to submit their solutions multiple times with the selected tests being run each time
and feedback given to the student. 
We further note that in typical algorithms courses,  testing inputs must be large enough to ensure that the students' solutions have 
the required asymptotic  complexity. 
In such scenarios, course instructors typically resort to semi-random test generation, selecting only a small  number of the generated
tests due to the limited computational resources of testing platforms. 
Hence, the included tests must be judiciously chosen. 
Manual strategies for test selection, however, do not scale for large testing inputs.

This paper presents \testselector, a new framework for optimal test selection for student projects. 
With our framework, the instructor provides a canonical implementation of the project 
assignment, a set of generated tests $T$, and the number $n$ of tests to be selected, and \testselector 
determines a subset $T' \subseteq T$  of size $n$ that maximises a given code coverage measure. 
By maximising coverage of the canonical solution, \testselector provides relative assurances that most 
of the corner case behaviours of the expected solution are covered by the selected test suite.
Naturally, the better the coverage measure, the better those assurances. 
Importantly, the best coverage measure is often project-specific, there being no silver bullet. 

The key advantage of \testselector over existing approaches~\cite{yamada:icst:2015,hnich:constraints:2006,chen:seke:2008,kitamura:safecomp:2018} is precisely that it is easily extensible with arbitrarily 
complex code coverage measures specifically designed for the project at hand. 
Unlike previous approaches however, \testselector does not require the targeted coverage measures to be encoded into the logic of an 
exact constraint solver. 
We achieve this by using as our optimisation algorithm, a specialised version of the recent \seesaw algorithm~\cite{seesaw-cp21}
for exploring the Pareto optimal frontier of a pair of functions.  
We demonstrate the flexibility of \testselector by extending it with support for a 
range of classical code coverage measures and using it to select test suites for a number of real-world 
algorithms projects, further showing that the selected test suites outperform randomly selected ones in 
finding bugs in students' code.

The paper is organized as follows.
Section~\ref{sec:overview} overviews the \testselector framework presenting each of its modules and how they interact with each other.
Section~\ref{sec:instrumentation} and Section~\ref{sec:maxtests} describe in detail the main modules of \testselector. Section~\ref{sec:evaluation} presents an experimental evaluation of the framework.
Section~\ref{sec:related-work} overviews related work, and Section~\ref{sec:conclusion} concludes the paper.
 \section{TestSelector Overview}\label{sec:overview}

\newcommand{\etag}[1]{{\small \bf (#1)}}

\begin{figure}[t!]
  \centering
  \includegraphics[width=0.8\textwidth]{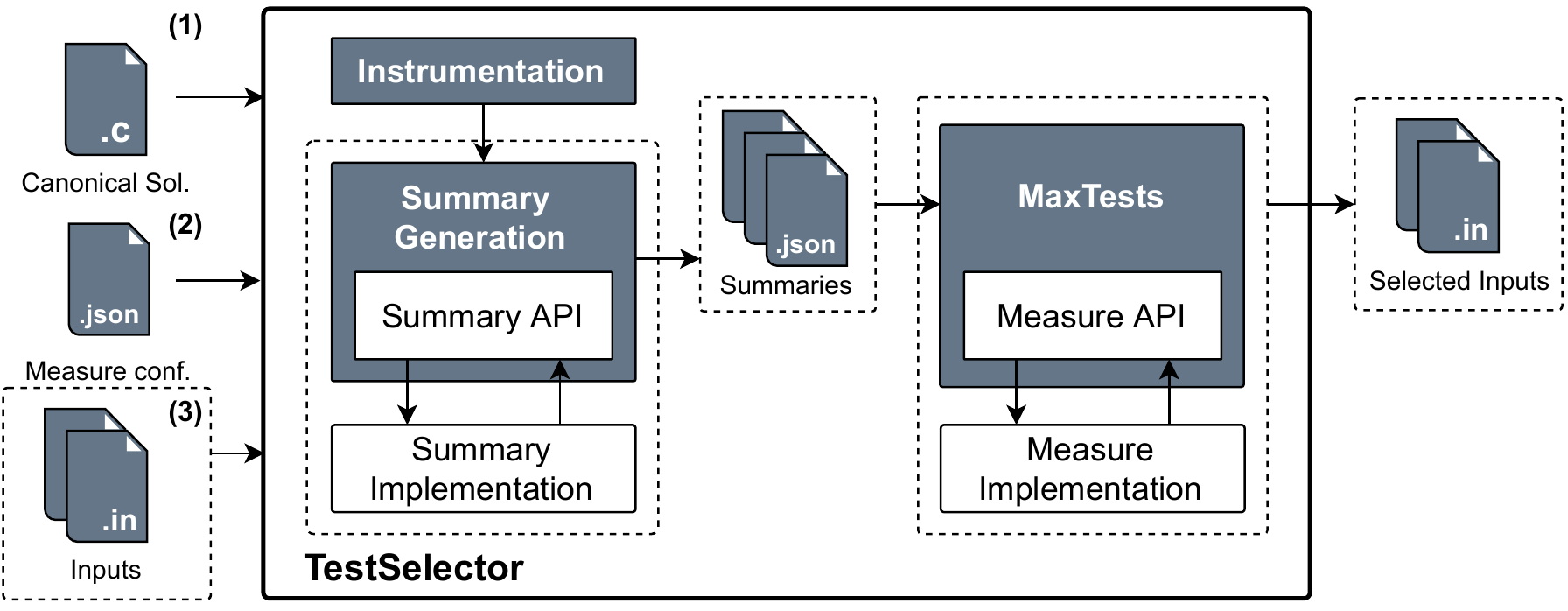}
  \caption{\testselector high-level architecture.}\label{fig:architecture}
\end{figure}

We give an overview of our approach for selecting optimal test suites for 
student projects. 
As illustrated in Figure~\ref{fig:architecture}, the \testselector framework receives three inputs: 
\etag{1} the instructor's implementation for the project, which we refer to as the \emph{canonical solution}; 
\etag{2} a JSON configuration file with a description of the coverage measure to be used for test selection as well as the number of tests to be selected; 
and \etag{3} an initial set of input tests, $T$. 
Given these inputs, \testselector computes an optimal subset of tests, $T' \subseteq T$, that maximises 
the selected coverage measure with a fixed number of tests, $n$ ($\vert T' \vert = n$). 
Due to the combinatorial nature of the problem and the sheer size of the search space, it is often the case that  
\testselector is not able to find the optimal solution within the given time constraints.
In such cases, it returns the best solution found so far. 
Our experimental evaluation indicates that this solution is typically not far from the optimal one. 

 The \testselector framework consists of two main building blocks: 
\begin{itemize}
  \item \emph{Summary Generation Module:} The summary generation module automatically instruments
  the code of the canonical solution in order for its execution to additionally produce a \emph{coverage summary} of 
  each given input test. 
Different coverage measures require different summaries. 
For instance, a \emph{block coverage summary} (c.f. \S\ref{subsec:measures})
  simply includes the identifiers of the code blocks that were executed during the running of the canonical solution. 
\item \emph{\maxtests Module:} The \maxtests module receives as input 
   the coverage measure to be used, the number $n$ of tests to be selected, and a set of summaries, and selects 
   the subset of size $n$ of the given summaries that maximises the coverage measure. For instance, 
   for the \emph{block coverage measure}, \maxtests selects  the summaries corresponding to the testing inputs that maximise 
   the overall number of executed code blocks. 
Note that if \maxtests does not find the optimal solution within 
   the specified time limit, it simply outputs the best solution found so far.  
   
   At the core of \maxtests is an adapted implementation of the \seesaw algorithm~\cite{seesaw-cp21}, a novel
   algorithm for exploring the Pareto optimal frontier of two given functions using the well-known implicit hitting set paradigm~\cite{DaviesB11,DaviesB13}.
   The key innovation of \seesaw is that it allows one to treat one of the two functions to optimise in a black-box manner. 
   In our case, this black-box function corresponds to the targeted coverage function, meaning that we are able to select
   optimal test suites without encoding the targeted coverage functions into the logic of an exact constraint solver. 
   
\end{itemize}

\subsection{Supporting New Coverage Measures}
\label{subsec:extensiblility}

The key advantage of \testselector when compared to existing approaches for constraint-base test suite selection~\cite{yamada:icst:2015,hnich:constraints:2006,huayao:corr:2019,chen:seke:2008,kitamura:safecomp:2018} 
is that it is trivial to extend \testselector with support for new, arbitrarily complex coverage measures.
In contrast, existing approaches require users to encode the targeted coverage measures into the logic 
of an exact constraint solver, typically SMT~\cite{de_moura:2008} or Integer Linear Programming (ILP) solvers~\cite{gurobi}.
The manual construction of such encodings has two main inconveniences when compared to our approach. 
First, it requires requires specialist knowledge on the logic and inner workings of the targeted solver. 
Note that even simple encodings must be carefully engineered so that they can be efficiently solved. 
Second, there might be a mismatch between the expressivity of the existing solvers and the nature 
of the measure to be encoded. 
In contrast, with \testselector, if one wants to add support for a new coverage measure, one simply has to: 
\begin{enumerate}
  \item Implement a \emph{Coverage Summary API} that dynamically constructs a coverage summary 
            during the execution of the canonical solution;
  
\item Implement a \emph{Coverage Evaluation Function} that maps a given set of coverage summaries to a
           numeric coverage score.  Importantly, in order for \testselector to work properly, the coverage evaluation function 
           must be \emph{monotone}; meaning that for any two sets 
           of summaries $S_1$ and $S_2$, it must hold that: 
           $S_1 \subseteq S_2 \implies f(S_1) \leq f(S_2)$.
           Monotonicity is a natural requirement for coverage scoring functions. Hence, we do not believe that 
           this restriction constitutes a limitation to the applicability of our framework. 
\end{enumerate}

\subsection{Natively Supported Coverage Measures}
\label{subsec:measures}

Even though our main goal is to allow for users to easily implement their own coverage measures,
\testselector comes with built-in support for various standard code coverage measures. In particular, 
it implements\footnote{Note that we use $\#X$ to refer to the number of elements of $X$.}: 
\begin{itemize}
\item \emph{Block Coverage (BC)} --- counts the number of executed code blocks: $$
 f_{BC}(S) = \#\{ i \mid \exists s \in S. \ s \ \text{contains an execution of } i \} 
$$

\item \emph{Array Coverage (AC)} --- counts the number of programmatic interactions with distinct array indexes: 
$$
 f_{AC}(S) = \#\{ (a, i) \mid \exists s \in S. \ s \ \text{contains an access to the $i$-th position of $a$}   \} 
$$ 
\item \emph{Loop Coverage (LC)} --- counts the number of loop executions with a distinct number
of iterations: $$
f_{LC}(S) = \#\{ (l, i) \mid  \exists s \in S. \ s \ \text{contains an execution of loop $l$ with $i$ iterations} \}
$$
\item \emph{Decision Coverage (DC)} --- counts the number of conditional guards that evaluate both
to \textbf{true} and to \textbf{false}: $$
f_{DC}(S) = \#\left\{ 
        {\begin{array}{l}
           i \mid  \exists s_1, s_2 \in S. \ s_1  \ \text{evaluates the $i$-th guard to } \textbf{true}  \\ 
             \quad \wedge \ s_2  \ \text{evaluates the $i$-th guard to }  \textbf{false}
        \end{array}}
        \right\}
$$

\item \emph{Condition Coverage (CC)} --- counts the number of conditional guards for 
which all subexpressions evaluate both to \textbf{true} and to \textbf{false}: $$
f_{CC}(S) = \#\left\{ 
        {\begin{array}{l}
           i \mid  \exists s_1, s_1', ..., s_n, s_n' \in S.  \ \forall 1 \leq j \leq n .   \\ 
             \ s_j \text{ evaluates the $j$-th sub-expr. of the $i$-th guard to } \textbf{true} \\ 
             \ s_j' \text{ evaluates the $j$-th sub-expr. of the $i$-th guard to } \textbf{false} \\ 
        \end{array}}
        \right\}
$$

\end{itemize}
We refer the reader to~\cite{Szugyi:2013} for a detailed account of standard coverage measures 
in the software engineering literature.

\paragraph{Linear Combination of Coverage Measures.}
In addition to the coverage measures described above, \testselector allows the user to specify 
a linear combination of coverage measures. 
Observe that, as the linear combination of two monotone functions is also monotone, 
the user is free to combine any monotone coverage measures without 
compromising the correct behaviour of \maxtests.

 \section{Summary Generation}\label{sec:instrumentation}

\begin{figure}[h!]
\centering
\includegraphics[width=0.8\textwidth]{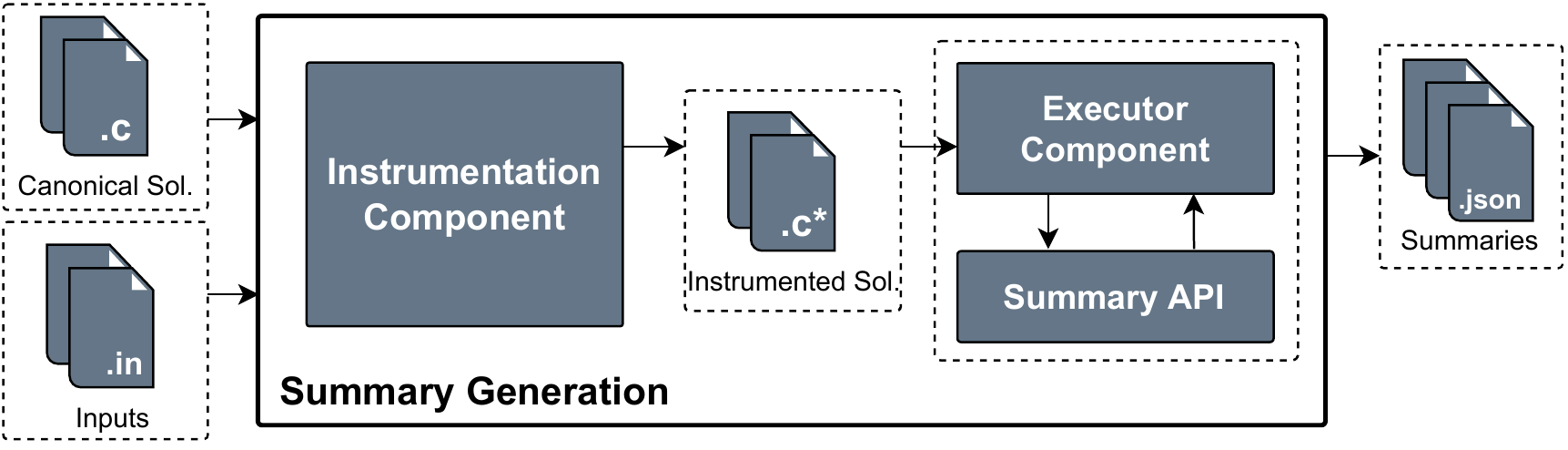}
  \caption{Summary Generation Module: Architecture.}\label{fig:summ-gen}
\end{figure}

This section overviews the \emph{Summary Generation Module} of \testselector
which, given a canonical solution and a set of testing inputs, generates the corresponding 
set of summaries with the relevant coverage data. 
The internal architecture of the module, described in Figure~\ref{fig:summ-gen}, 
comprises two components: 
\begin{itemize}
  \item the \emph{instrumentation component}, described in \S\ref{subsec:instrumentation}, which injects into the code 
          of the canonical solution calls to the coverage summary API before and/or after each summary-relevant operation;
\item the \emph{executor component}, described in \S\ref{subset:sum:gen}, which runs the instrumented code of the canonical solution 
            on the given set of testing inputs using the  appropriate implementation(s) of the coverage summary API\@. 
\end{itemize} 

\subsection{Program Instrumentation}
\label{subsec:instrumentation}

\newcommand{\code}[1]{\texttt{#1}}

The main job of the program instrumentation component is to inject calls to the coverage summary 
API into the code of the canonical solution. 
For instance, given the program: 
\begin{minted}[fontsize=\footnotesize]{c}
while ((x--) > 0) { aux = a[0]; a[0] = aux + x; }
\end{minted}
The instrumentation component generates the program: 
\begin{listing}[H]
  \vspace{-10pt}
  \begin{minted}[fontsize=\footnotesize,gobble=4]{c}
    BEGIN("WHILE", 3);
    while (GUARD(COND((x--) > 0))) { 
      BLOCK(2);
      aux = a[0]; ARR_READ("main:a", a, 0);
      a[0] = aux + x; ARR_WRITE("main:a", a, 0);
    }
    END("WHILE", 3);
  \end{minted}
  \vspace{-5pt}
  \caption{Instrumented program.}\label{lst:instr-prog} 
  \vspace{-5pt}
\end{listing}

\vspace{-10pt}
The example above showcases the API functions: \code{BEGIN}, \code{END}, \code{BLOCK}, \code{GUARD},  
\code{COND}, \code{ARR\_READ}, and \code{ARR\_WRITE}. 
In a nutshell, we inject a call to \texttt{BEGIN} and \texttt{END} respectively before and 
after each control-flow statement, providing both the type of control-flow statement and its unique 
identifier. 
We inject a call to \code{BLOCK} at the beginning of each conditional/loop branch. 
 All conditional guards are wrapped inside a call to the API function \code{GUARD} and 
all Boolean sub-expressions of a guard are wrapped inside a call to \code{COND}. 
The instrumentation also ensures that all array-lookup and array-update operations are respectively succeeded by a call to \code{ARR\_READ} and \code{ARR\_WRITE}, 
providing the static identifier of the array, the pointer to the array, and the accessed index. 
Subsection \S\ref{subset:sum:gen} gives a more comprehensive account of the coverage summary API, 
while below we discuss the instrumentation procedure.

\begin{figure}[t!]
  \begin{mathpar}
  \inferrule[If]{s_i' = \texttt{BLOCK}(\getId(s_i)); \compiler{s_i, \getId(s_i)}\mid_{i=1, 2}
    \\\\ 
    e' = \ecompiler{e}
  }{{\begin{array}{l}
    \compiler{\ifstmt{e}{s_1}{s_2}, i} \semeq \\ 
      \quad\left\{
        \begin{array}{l}
          \mathtt{BEGIN}(\mintinline{c}{"IF"}, i); \\
          \mintinline{c}{if}\,(\mathtt{GUARD}(e'))\,\{\stmt_1'\}\,
          \mintinline{c}{else}\,\{\stmt_2'\} \\
          \mathtt{END}(\mintinline{c}{"IF"}, i);
        \end{array}
      \right.
    \end{array}}
  }
  \and
  \inferrule[While]{\stmt' = \getId(s_i); \compiler{\stmt, \getId(s)}
    \\\\ 
    e' = \ecompiler{e}
  }{{\begin{array}{l}
      \compiler{\whilestmt{e}{s}, i} \semeq \\ 
      \quad\left\{
        \begin{array}{l}
          \mathtt{BEGIN}(\mintinline{c}{"WHILE"}, i); \\
          \mintinline{c}{while}\,(\mathtt{GUARD}(e'))\,\{\stmt' \} \\
        \mathtt{END}(\mintinline{c}{"WHILE"}, i); \\
        \end{array}
      \right.
    \end{array}}
  }
  \and
    \inferrule[Sequence]{}{\compiler{s_1; s_2, -} \semeq \compiler{s_1, \getId(s_1)}; \compiler{s_2, \getId(s_2)}}
  \and
  \inferrule[Array Write]{\idx = \freshV()  
      \and 
      \str_x = \mathtt{string}(x)
      }{{\begin{array}{l}
      \compiler{x\left[e_1\right] = e_2} \semeq
      \left\{
        \begin{array}{l}
          \mintinline{c}{int } \idx  = e_1; \\
           x\left[\idx\right] = e_2; \\
          \mathtt{ARR\_WRITE}
            \left(
              \str_x, 
              (\mintinline{c}{void *})\ x,\ \idx
            \right); 
        \end{array}
      \right.
    \end{array}}
  }
  \\\\
    \inferrule[Malloc]{\arg = \freshV()
  }{{\begin{array}{l}
      \compiler{x = \mathtt{malloc}\ (e)} \semeq \\
      \quad\left\{
        \begin{array}{l}
          \mintinline{c}{int } \arg = e; \\
          \mathtt{CALL\_MALLOC(\arg)}; \\
          x = \mathtt{malloc}(\arg);
        \end{array}
      \right.
    \end{array}}
  }
  \and
  \inferrule[Array Read]{\idx = \freshV()  
      \and 
      \str_x = \mathtt{string}(x)
      \and 
      \tau = {\sf Type}(e_1)
  }{{\begin{array}{l}
     \compiler{x = e_1\left[e_2\right]} \semeq \\
      \quad\left\{
        \begin{array}{l}
          \tau \ \auxV = e_1; \ \mintinline{c}{int } \idx = e_2; \\
          x = \auxV[\idx]; \\
          \mathtt{ARR\_READ}
            \left(
              \str_x, \
              (\mintinline{c}{void *})\ \auxV,\
              \idx
            \right); 
        \end{array}
      \right.
    \end{array}}
  }
  \end{mathpar}
  \caption{Instrumentation function $\compfun : \stmts \times \ints \rightarrow \stmts$.}
  \label{inst:function}
\end{figure}

\paragraph{Formalisation.}
We formalise our instrumentation procedure for the fragment of the C programming language
given below. Note that the implementation of \testselector supports the entire syntax of C. 

{\small 
\begin{equation*}
  \begin{tabular}{rl}
    $\stmt \in \stmts$ ::= & $
     \ifstmt{e}{\stmt_1}{\stmt_2} \ \big| \
     \whilestmt{e}{\stmt} \ \big| \ 
     s_1; s_2 \ \big| \ 
     x = e$ \\ 
& $\ \big| \
     x[e_1] = e_2 \ \big| \  
     x = e_1[e_2] \ \big| \ 
     x =  \mathtt{malloc}(e) 
    $ \\[5pt]
$e \in \exprs$ ::= & $v \ \big| \ x \ \big| \ \ominus e \ \big| \ e_1 \oplus e_2 $
  \end{tabular}
\end{equation*}}

\noindent Statements $\stmt \in \stmts$ include the standard conditional, loop, and 
sequence statements, variable assignments, and array updates, lookups, and creations. 
Statements are composed of expressions, $e \in \exprs$, which include: constants 
$v \in \vals$, program variables $x \in \vars$, and unary and  binary operators. 
In the following, we assume that each statement $s$ is annotated with a unique integer identifier,
denoted by $\getId(s)$.
 This can be achieved via a simple traversal of the AST of the program to be instrumented. 
We formalise the instrumentation as a total compilation function $\compfun : \stmts \times \ints \rightarrow \stmts$ 
mapping pairs of statements and integer identifiers to statements and write $\compfun(s, i)$ to denote the instrumentation of 
the statement $s$ with identifier $i$. 
The compilation rules are given in Figure~\ref{inst:function} and are mostly straightforward. 
We inject calls to the coverage summary  API before and/or after  each summary-relevant operation. 
For array-manipulating operations, we have to store the computed index (and pointer) to avoid 
re-computation and repetition of side-effects. 
Finally, our main instrumentation function makes use of an auxiliary instrumentation function 
for expressions $\mathcal{C}_e : \exprs \mapsto \exprs$, which simply wraps all the Boolean 
sub-expressions of the given expression within a call to the 
\texttt{COND} API function. Below, we illustrate the two cases corresponding to binary operator 
expressions: \begin{mathpar}
\inferrule[BinOp - COND]{
\isBool(\oplus) 
}{
  \ecompiler{e_1 \oplus e_2} = \mathtt{COND}(\ecompiler{e_1}) \oplus \mathtt{COND}(\ecompiler{e_2})   
}
\and 
\inferrule[BinOp - Non-COND]{
\neg \isBool(\oplus) 
}{
  \ecompiler{e_1 \oplus e_2} = \ecompiler{e_1} \oplus \ecompiler{e_2}
}
\end{mathpar}

\subsection{Summary Generation API}\label{subset:sum:gen}

\begin{figure}[b!]
  \begin{tabular}{lll}
    \mintinline{c}{BEGIN(char *stmt_t, int id)} & & \begin{tabular}[c]{@{}l@{}}
      Logs the beginning of a control-flow statement with \\
      type \texttt{stmt\_t} and identifier \texttt{id}.
    \end{tabular} \\[10pt]
\mintinline{c}{END(char *stmt_t, int id)} & & \begin{tabular}[c]{@{}l@{}}
      Logs the ending of a control-flow statement with \\
     type \texttt{stmt\_t} and identifier \texttt{id}.
    \end{tabular} \\[10pt]
\mintinline{c}{BLOCK(int id)} & & \begin{tabular}[c]{@{}l@{}}
      Logs the block identifier \texttt{id}.
    \end{tabular} \\[5pt]
\mintinline{c}{GUARD(int expr)} & & \begin{tabular}[c]{@{}l@{}}
      Logs the value \texttt{expr} of a conditional guard.
    \end{tabular} \\[5pt]
\mintinline{c}{COND(int expr)} & & \begin{tabular}[c]{@{}l@{}}
      Logs the value \texttt{expr} of a boolean sub-expression \\ of conditional guard.
    \end{tabular} \\[10pt]
\begin{tabular}[c]{@{}r@{}}
      \mintinline{c}{ARR_WRITE(char *name,} \\
      \mintinline{c}{void *ptr, } \\
      \mintinline{c}{int i)     }
    \end{tabular} & & \begin{tabular}[c]{@{}l@{}}
      Logs an update operation to the index \texttt{i} of the \\ 
      array pointed to by \texttt{ptr} and associated with \\ 
      the static identifier \texttt{name}. 
    \end{tabular} \\[15pt]
\begin{tabular}[c]{@{}r@{}}
      \mintinline{c}{ARR_READ(char *name,} \\
      \mintinline{c}{void *ptr, } \\
      \mintinline{c}{int i)     }
    \end{tabular} & & \begin{tabular}[c]{@{}l@{}}
      Logs a lookup operation of the index \texttt{i} of the \\ 
      array pointed to by \texttt{ptr} and associated with \\ 
      the static identifier \texttt{name}. 
    \end{tabular} \\[15pt]
\mintinline{c}{CALL_MALLOC(unsigned size)} & & \begin{tabular}[c]{@{}l@{}}
    Logs a call to \texttt{malloc} with argument \texttt{size}.
    \end{tabular} \\
  \end{tabular}
  \caption{Coverage Summary API (selection).}\label{fig:api}
\end{figure}

Figure~\ref{fig:api} gives a selection of the functions that form our coverage summary API;
the description of the full API can be found in the appendix. 
The purpose of the API is to dynamically compute a coverage summary for each testing input. 
Note that different coverage measures require different summaries and therefore different 
implementations of the coverage summary  API.  
Instead, we could have opted to serialise the entire execution trace and then compute 
the coverage function directly on the serialised trace. 
This approach, however, does not scale to executions with millions of commands required 
to assess typical algorithms projects. 
For this reason, each coverage measure is associated with a particular implementation 
of the coverage summary API. 
For portability, \testselector requires coverage summaries to be serialised as JSON documents.

Let us now take a closer look at the implementation of the coverage summary API. 
We will specifically consider the implementation of the API for the block coverage, array coverage, and loop coverage measures,
as the implementations of the other measures are analogous. 
We will again use as an example the program:
\begin{minted}[fontsize=\footnotesize]{c}
while ((x--) > 0) { aux = a[0]; a[0] = aux + x; }
\end{minted} 
whose instrumentation was given in Listing~\ref{lst:instr-prog}, and will assume that 
it is run with the testing input $\mathtt{x} = 4$.

\paragraph{Block Coverage.} 
The block coverage summary generated for the example is \mintinline{json}|{ "2" : true }|, simply stating that the code block with identifier $2$ was covered 
 by the testing input $\mathtt{x} = 4$. 
The API implementation for the block coverage measure is straightforward:
the function \mintinline{c}{BLOCK(int id)} has to set the summary entry 
corresponding to \texttt{id} to \mintinline{c}{true} and
all other API functions can be ignored.\footnote{We provide default implementations to be used when no summary-related 
behaviour should be triggered.} 

\paragraph{Array Coverage.} 
The array coverage summary generated for the example is:
\begin{minted}[fontsize=\footnotesize]{json}
{ "main_a" : { "read" : [0], "write" : [0] } }
\end{minted} 
signifying that there was a lookup operation and an update operation at the index $0$ of the array with 
static identifier  \mintinline{c}{a} in function \texttt{main}.   
The API implementation for the array coverage measure is also straightforward:
the functions \texttt{ARR\_READ} and \texttt{ARR\_WRITE} have to add the inspected index
to the read and write lists of the corresponding array.

\paragraph{Loop Coverage.} 
The loop coverage summary generated for the example is:
\begin{minted}[fontsize=\footnotesize]{json}
{ "3" : { "4" : 1 } }
\end{minted} 
signifying that the loop with identifier $3$ had one execution with $4$ iterations. 
The API implementation for the loop coverage measure is a bit more involved than the previous ones. 
We use the \texttt{BEGIN} and \texttt{GUARD} functions to count the number of consecutive 
iterations of a given loop and then update the corresponding summary entry when the 
corresponding \texttt{END} function is called. To this end, we have to maintain a stack of 
counters, each corresponding to an active loop.

 \section{MaxTests: Constraint-based Test Suite Selection}\label{sec:maxtests}

\newcommand{\tests}[0]{\mathcal{T}}
\newcommand{\sums}[0]{\mathcal{S}}

\begin{figure}[h]
  \centering
  \includegraphics[width=0.60\textwidth]{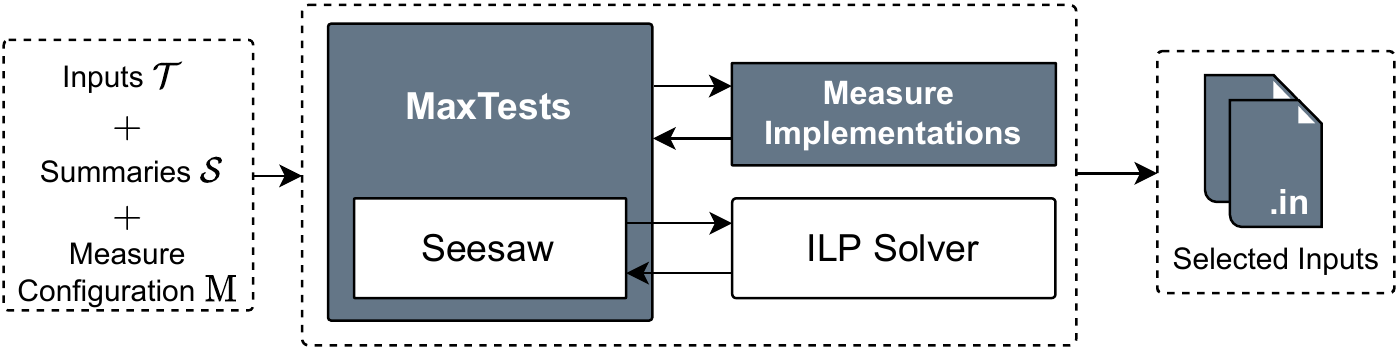}
  \caption{MaxTests Module: Architecture.}\label{fig:maxtests-arch}
\end{figure}

The \maxtests module is responsible for computing the optimal set of testing inputs according to the chosen measure.
This module, whose architecture is presented in Figure~\ref{fig:maxtests-arch}, is composed of two main components: 
\begin{itemize}
\item the \emph{Seesaw Component:} a specialisation of the Seesaw algorithm~\cite{seesaw-cp21} for the test selection problem; and
\item the \emph{Measure Component:} the implementation of the targeted measure function, which receives as input a set of summaries and generates the corresponding 
         numeric coverage score. 
\end{itemize}
Given a set of testing inputs $\tests$, their corresponding summaries $\sums$, the number~$n$ of testing inputs
to select, and the targeted coverage measure $M$, \maxtests determines a subset $T \subseteq \tests$ of size $n$ that maximises
the specified measure. 

As we have seen before, coverage measures are not computed directly on the testing inputs but rather on their 
corresponding summaries. 
Hence, one can think of a coverage measure, $M : 2^{\summaries} \rightarrow\reals$, as a mathematical function mapping 
sets of summaries to their corresponding coverage score. 
Accordingly, in order to find the optimal set of tests $T \subseteq \tests$, \maxtests first determines the optimal subset 
 of summaries $S \subseteq \sums$ and then outputs their corresponding testing inputs. 

If one wants to extend \testselector with support for a new coverage measure, one need only provide 
its corresponding implementation (i.e., the function~$M$), with the code of \seesaw not requiring any adaptation. 
The \seesaw algorithm requires a back-end ILP solver. Here, we use Gurobi~\cite{gurobi}; however, \maxtests is structured so that it is 
trivial to replace Gurobi with any other ILP~solver. 
In the following, we explain our specialisation of \seesaw to the test selection problem, 
highlighting domain-specific choices and design decisions.

\paragraph{Specialised Seesaw Algorithm for Test Selection.}
The \seesaw algorithm \cite{seesaw-cp21} was designed to explore the Pareto optimal frontier of two functions: 
\etag{1} a \emph{cost} function $g$, that must be encodable into the logic of an exact solver, such as an ILP solver (e.g.~\cite{gurobi}) or a MaxSAT solver (for e.g.~\cite{DaviesB13}), and
\etag{2} an \emph{oracle} function $f$, that is treated in a black-box fashion and only required to be monotone.  
In the case of the test selection problem, the oracle function $f$ corresponds to the coverage measure $M$, while the 
cost function $g$ is simply a Boolean function indicating whether or not the cardinality of the set of selected tests is equal to the required number of tests. 
Our goal is to maximise both functions: we want the maximum coverage with the specified number of  
testing inputs.

\newcommand{\sbest} {S_{best}}
\begin{algorithm}[h]
\caption{Specialised \seesaw Algorithm.}\label{algorithm:maxtests}
  
  \Input{$\ntests$, $\measure$, $\summaries$\tcp*[r]{$\measure\;:\; 2^{\summaries} \rightarrow\reals$}}
  \Output{subset of $\summaries$ with size $\ntests$ that maximises $\measure$}
  
  \BlankLine
  
  $S_{best}\gets\maxheuristic(\measure, \summaries, \ntests)$\tcp*[r]{heuristic candidate of size $\ntests$}
  
  $\ABSTRACTION\gets \emptyset$\tcp*[r]{set of collected cores}
  
  \While{\true} {\(S\gets\text{arg}_{S\in\hittingsets(\ABSTRACTION)} |S| = \ntests\)\tcp*[r]{find minimal  hitting set of size $\ntests$}

    \If{\(S =\bot\)\label{algorithm:maxtests:nomore}} {
      \Return $\sbest$
    }
    
    \If{$\measure(S) = \measure(\summaries)$ \label{algorithm:maxtests:maxcov}} {
      \Return $S$
    }
    
    \If(\tcp*[f]{ upper bound bound improvement}){$\measure(S) > \measure(\sbest)$\label{algorithm:maxtests:improve}} {
        $\sbest \gets S$
    }
    
    $\ABSTRACTION\gets\ABSTRACTION\cup\{\getCore(S, \measure, \summaries)\}\label{algorithm:maxtests:extractcore}$\tcp*[r]{calculate new  core}
  }
\end{algorithm}

Algorithm~\ref{algorithm:maxtests} presents the pseudo-code of \seesaw adapted to the test selection problem.
The algorithm goes through a sequence of sets of size $\ntests$ called \emph{candidates}, starting with an initial candidate that is heuristically determined. At each iteration, the algorithm tries to compute a new candidate with a better coverage score. 
To do so, it enumerates necessary conditions for coverage improvement, one by one.
These conditions take the form of sets of summaries called \emph{cores}. 
In a nutshell, in order to improve the coverage score of the current best solution, one has to find another solution that includes at least one summary belonging to each computed core.  
All computed cores are accumulated in the variable $\ABSTRACTION$, and the new candidates are chosen to be hitting sets of $\ABSTRACTION$ of size $\ntests$,\footnote{This 
step of the algorithm is computed with the help of the underlying ILP solver (in our case, Gurobi).} meaning that they must include at least one 
summary belonging to each core in $\ABSTRACTION$. 
If no more candidates exist, that is, $S =\bot$ in line \ref{algorithm:maxtests:nomore}, then the best candidate so far is returned.
In line~\ref{algorithm:maxtests:maxcov}, if the coverage score of the current candidate is equal to the maximum possible coverage (i.e. the coverage of the entire set of summaries), then the algorithm also outputs the current candidate.
In line~\ref{algorithm:maxtests:improve}, if the coverage score of the current candidate improves on the coverage of current best candidate, then the best candidate $\sbest$ is updated.
Finally, in line~\ref{algorithm:maxtests:extractcore}, the algorithm extends the set of computed cores, $\ABSTRACTION$, with a new core computed using the current candidate.  
Below, we detail this process.

\paragraph{Obtaining New Cores from Candidates.}
\label{subsec:extractcore}

New cores are obtained using the auxiliary function $\getCore$.
For a candidate $S$ with a coverage value $\nu$ ($\nu = \measure(S)$), the computation of core extraction is processed in two steps:
\begin{enumerate}
\item Choose a subset-maximal set $S'$ such that $S \subseteq S'$ and $\measure(S') = \nu$;
\item Return the core $\sums \setminus S'$.
\end{enumerate}
The intuition is simple: if $S'$ is larger than $S$ and does not have a higher coverage score; then,
in order to improve the coverage score of $S$, one has to pick a summary in $\sums \setminus S'$.  
Naturally, the smaller the chosen cores, the faster the algorithm converges to an optimal 
solution. The question is how to efficiently find the largest possible $S'$ (corresponding to the smallest possible core). 
Here we consider two heuristics for core computation: \emph{linear search} and \emph{progression~search}.

Algorithm~\ref{algorithm:lsecore} presents the pseudo-code of the linear search $\getCore$ function.
We start by ordering all summaries according to their individual coverage score  (in increasing order). Then, we traverse the array of ordered summaries, checking if the inclusion of each summary that does not belong to the current candidate maintains the coverage score of the candidate. 
If it does not, then we discard that test; otherwise, we add it to the current candidate.

\begin{algorithm}[h]
\caption{Linear Search $\getCore$}\label{algorithm:lsecore}
  
  \Input{$S$, $\measure$, $\summaries$\tcp*[r]{$\measure\;:\; 2^{\sums}\rightarrow\reals$}}
  \Output{$\kappa$\tcp*[r]{a core}}
  
  \BlankLine
  
  $\sums_{ord}\gets \order(\measure, \summaries)$\label{algorithm:extractcore:ord}\tcp*[r]{summaries in increasing measure order}
  
  $S' \gets S$\;
  
  \ForEach{$s \in \sums_{ord}$ \and $ s\notin S'$\label{algorithm:extractcore:loop}} {

    \If{$\measure(S'\cup \{s\}) = \measure(S')$ \label{algorithm:extractcore:condition}} {
      $S' \gets S' \cup \{s\}$
    }
    
  }
  \Return $\sums\setminus S'$
\end{algorithm}

The linear search strategy has one major disadvantage: it requires re-comput-ing the coverage measure for each summary 
that does not belong to the current candidate. 
The goal of the progression search strategy~\cite{progressionecai14} is precisely to mitigate this problem by reducing 
the number of calls to the measure function.
As summaries are ordered in an array-like fashion (line~\ref{algorithm:extractcore:ord}), instead of considering one summary at 
a time (as in lines~\ref{algorithm:extractcore:loop} to~\ref{algorithm:extractcore:condition}), we consider summaries in 
exponential progression, that is, we consider extending the current candidate $S'$ first with 1 summary, then with 2, then with 4, and so on.
If the coverage score of $S'$ with the additional summaries is equal to the coverage score of $S$, then we add the summaries to the current candidate $S'$; 
otherwise, we restart the progression from the last reached point.

 \section{Evaluation}\label{sec:evaluation}

We evaluate \testselector with respect to three research questions:
\begin{itemize}
  \item \textbf{RQ1: How easy is it to extend \testselector with new code coverage measures?} We show that the currently supported coverage measures are implemented with a small number of lines of code, demonstrating the practicality of our approach.
  \item \textbf{RQ2: Do classical code coverage measures improve test suite selection for bug finding in student projects?} We show that the test suites selected by \testselector outperform randomly selected ones in finding bugs in students' code.
  \item \textbf{RQ3: Do linear combinations of code coverage measures further improve test suite selection for bug finding?} We show that by combining the best code coverage measures, we can find more bugs in students'~code.
\end{itemize}

\begin{figure}[t]
  \begin{minipage}[b]{0.49\textwidth}
    \centering
    \includegraphics[width=1.1\textwidth]{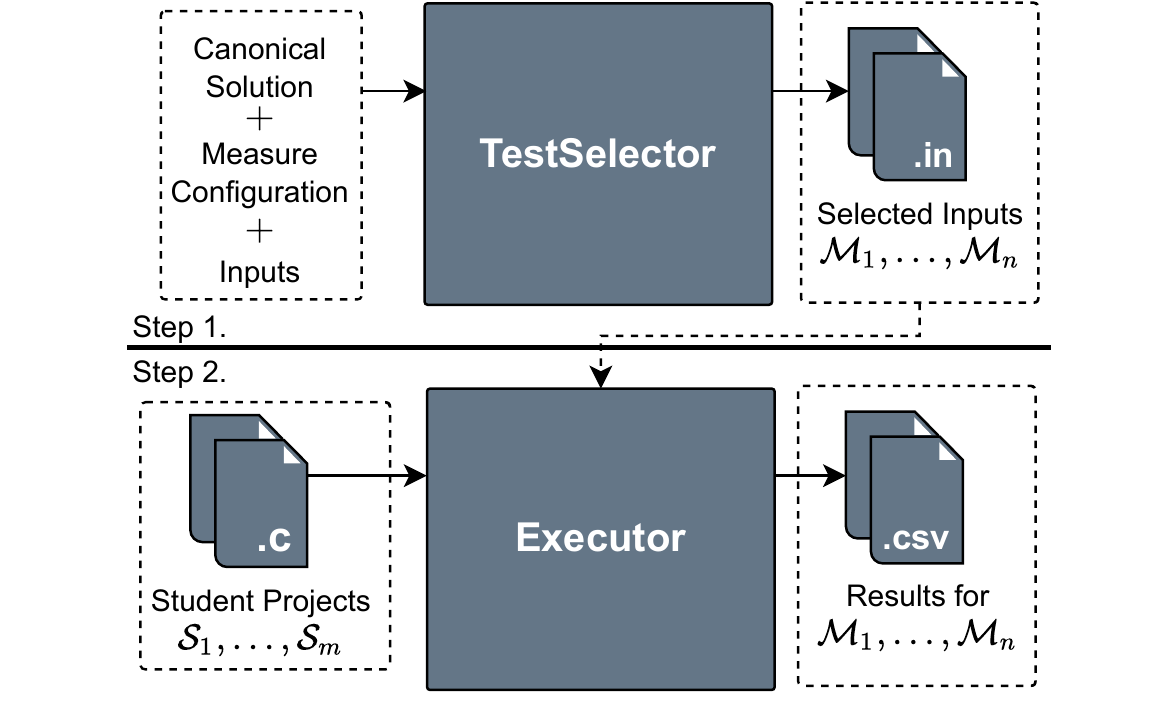}
    \captionof{figure}{Evaluation diagram.}\label{fig:eval-diag}
  \end{minipage}
  \hfill
  \begin{minipage}[b]{0.49\textwidth}
  \centering
  \resizebox{\linewidth}{!}{
  \begin{tabular}{crrrrr}
    \toprule
    Project & $\mathcal{C}_\text{LoC}$ & $n_\mathit{proj}$ & $\text{T}_\text{LoC}$ & $\text{Avg}_\text{LoC}$ & $n_\mathit{inpts}$\\
    \midrule
    P1 &   256 &   398 & 140,349 & 352.64 & 1,002\\
    P2 &   529 &   349 & 176,547 & 505.86 &   600\\
    P3 &   416 &   193 &  26,890 & 139.32 & 1,000\\
    P4 &   208 &   166 &  34,512 & 207.90 & 1,000\\
    P5 &   304 &   172 &  21,114 & 122.76 & 1,000\\
    P6 &   204 &   185 &  24,091 & 130.22 &   800\\
    P7 &   108 &   174 &  24,035 & 138.13 & 1,000\\
    \midrule
    Total        & 2,125 & 1,637 & 447,538 & 273.39 & 6,402\\
    \bottomrule
  \end{tabular}
  }
  \captionof{table}{Benchmark characterisation.}\label{table:benchmarks}
  \end{minipage}
\end{figure}

\paragraph{Experimental Procedure.} The experimental procedure is a two-step process, as illustrated in Figure~\ref{fig:eval-diag}. In the first step, \testselector selects the test suites for a given canonical solution, set of inputs, and configuration file specifying the coverage measures and the size of the computed test suites.
This step generates a set of test suites, each corresponding to one of the specified measures.
In the second step, an executor will run every student's project against the selected test suites. In the end, the executor creates a report detailing the passing/failing rate for every student's project on each selected test suite.\footnote{We consider that a student project fails a test if its output does not coincide with the output of the canonical solution or if it throws a runtime exception.}

\begin{figure}[t!]
\end{figure}

All the experiments were performed on a server with a 12-core Intel Xeon E5--2620 CPU and 32GB of RAM running Ubuntu 20.04.2 LTS. For the ILP solver we used the Gurobi Optimizer v9.1.2. For each execution of \maxtests we set a time limit of 30 minutes.

\paragraph{Benchmarks.} We curated a benchmark suite comprising students' projects from seven editions of two programming courses organised by the authors. Table~\ref{table:benchmarks} presents the benchmark suite characterisation. For each project, we show the number of lines of code of the canonical solution ($\mathcal{C}_\text{LoC}$), the number of student projects ($n_\mathit{proj}$), the total number of lines of code of the student projects ($\text{T}_\text{LoC}$), the average number of lines of code per student project ($\text{Avg}_\text{LoC}$), and the number of available input tests ($n_\mathit{inpts}$). In summary, we tested 1,637 projects, which totalled 447k lines of code ($\approx 273$ LoC/project).

\subsection{RQ1: \testselector Extensibility}

The table below presents the number of lines of code of the implementation of each coverage measure: \textit{Loop Coverage} (LC), \textit{Array Coverage} (AC), \textit{Block Coverage} (BC), \textit{Condition Coverage} (CC), and \textit{Decision Coverage} (DC). For each measure, we give the number of lines of code of both its implementation of the coverage summary API and evaluation function.

\begin{center}
\begin{tabular}{c ccccc}
    \toprule
    Module & LC & AC & BC & CC & DC \\
    \midrule 
    Coverage Summary API &  90 & 60 & 42 & 120 & 120 \\
    Measure Evaluation Function &  54 & 58 & 48 &  74 &  64 \\
    \bottomrule
\end{tabular}
\end{center}
When it comes to the implementation of the coverage summary API, we observe that the simpler coverage measures, such as LC, AC, and BC require fewer than 100 lines of code to implement and the more complex coverage measures, such as CC and DC, require 120 lines of code. As expected, the measure evaluation function is simpler to implement than the coverage summary API, requiring even fewer lines of code (between 48--74 LoC).

\subsection{RQ2: Classical Code Coverage Selection}\label{subsec:rq1}

We investigate the effectiveness of \testselector when used to select test suites for finding bugs in students' code. 
In particular, we compare the number of bugs found by the test suites selected by \testselector against those found by test suites obtained through random selection. 
In all experiments, we ask for test suites of size 30 out of 900 available randomly generated tests (the number of tests used to assess the students in the corresponding courses was~30).
We consider the five coverage measures described in \S\ref{subsec:measures} and an additional measure corresponding to the size of the testing input. 
Furthermore, to determine the best \textsf{extractCore} search strategy, \testselector was configured to run twice: one time using the linear search (LS) strategy 
and the other using the progression search (PS)~strategy.

\paragraph{Results.} Table~\ref{table:results-measures} presents the results of the experiment.
For each project, the table shows the resulting failure rates for the measures \textit{Loop Coverage} (LC), \textit{Array Coverage} (AC), \textit{Block Coverage} (BC), \textit{Size}, \textit{Condition Coverage} (CC), and \textit{Decision Coverage} (DC). We observe that the best measure is project-dependent, with LC being the best measure in four projects, BC in one, and Size in two. Importantly, we also observe that the more sophisticated measures, such as CC and DC, have lower failure rates than simpler measures, such as LC and AC\@. This may be explained by the fact that the students' most common programming errors are often encoded in loops and array accesses. Additionally, all coverage measures consistently perform better than the random test suite selection. 
Finally, we observe that the progression search strategy yields slightly better results than the linear search strategy.

\newcommand{\gf}[1]{\textcolor{gray}{\bf #1}}
\begin{table}[t]
  \centering
  \setlength{\tabcolsep}{7pt}
  \resizebox{\textwidth}{!}{
  \begin{tabular}{ccrrrrrrr}
    \toprule
    Project & Search & LC & AC & BC & Size & CC & DC & Rnd \\
    \midrule
    \multirow{2}{*}{P1}
                 &LS
                 &\bf{14.67}&14.48&13.69& 0.20&13.95&14.41& \multirow{2}{*}{4.81}  \\
                 &PS
                 &\gf{14.53}&14.34&13.56& 0.19&13.82&14.27&  \\
    \midrule
    \multirow{2}{*}{P2}
                 &LS
                 &18.07&17.15&{\bf19.47}& 6.14&15.60&14.35& \multirow{2}{*}{5.60} \\
                 &PS
                 &18.07&17.20&{\bf19.47}& 6.14&15.60&14.35& \\
    \midrule
    \multirow{2}{*}{P3}
                 &LS
                 &16.39&20.38& 7.49&\gf{28.07}& 7.49& 7.49& \multirow{2}{*}{7.56} \\
                 &PS
                 &16.77&20.70& 7.95&\bf{28.31}& 7.95& 7.95&  \\
    \midrule
    \multirow{2}{*}{P4}
                 &LS
                 &{\bf23.68}&22.78&11.99&23.59&17.95&17.93& \multirow{2}{*}{13.52} \\
                 &PS
                 &{\bf23.68}&22.82&11.99&23.59&17.93&17.93& \\
    \midrule
    \multirow{2}{*}{P5}
                 &LS
                 & {\bf3.76}& 3.23& 3.56& 3.74& 3.56& 3.56& \multirow{2}{*}{3.09} \\
                 &PS
                 & {\bf3.76}& 3.25& 3.56& 3.74& 3.56& 3.56&  \\
    \midrule
    \multirow{2}{*}{P6}
                 &LS
                 & 6.91& 8.22& 8.01&{\bf8.39}& 4.72& 4.68& \multirow{2}{*}{6.61} \\
                 &PS
                 & 6.91& 8.22& 8.01&{\bf8.39}& 4.72& 4.66&  \\
    \midrule
    \multirow{2}{*}{P7}
                 &LS
                 &{\bf10.46}& 6.08& 6.71& 6.39& 7.17& 7.17& \multirow{2}{*}{6.28} \\
                 &PS
                 &{\bf10.46}& 6.08& 6.71& 6.39& 7.17& 7.17&  \\
    \midrule
    \multirow{2}{*}{Average}
                 &LS
                 &\gf{13.42}&13.19&10.13&10.93&10.06& 9.94& \multirow{2}{*}{6.78} \\
                 &PS
                 &\bf{13.45}&13.23&10.18&10.96&10.11& 9.98& \\
    \bottomrule
  \end{tabular}
  }
  \caption{Results for each measure with linear search (LS) and progression search (PS).}\label{table:results-measures}
\end{table}

\begin{figure}[t]
  \centering
  \pgfplotstableread[col sep=comma,
                     search path=..]{tables/total_s2.csv}\tableone
  \pgfplotstableread[col sep=comma,
                     search path=..]{tables/total_s3.csv}\tabletwo
  \begin{tikzpicture}
    \begin{axis}[ybar=0.05cm,
        bar width = 5pt,
        height = 6cm,
        width = 8cm,
        ymin = 0,
        ymax = 15,
        scale only axis,
        axis lines = left,
        tick style={draw=none},
        enlarge x limits=0.1,
        xtick=data,
        xticklabels from table = {\tableone}{measure},
        x tick label style={rotate=45, anchor=east},
        y tick label style={
          /pgf/number format/.cd,
          fixed,
          fixed zerofill,
          precision=2,
          /tikz/.cd
        },
ylabel = {Failure Rate (\%)},
        ymajorgrids=true,
        grid style = dashed,
        legend image code/.code={\draw[#1] (0cm,-0.1cm) rectangle (0.15cm,0.2cm);
        },
        legend style = {
          anchor=north
        }
]
    \addplot table [
      x expr = \coordindex,
      y = {failure},
    ]{\tableone};
    \addplot table [
      x expr=\coordindex,
      y = {failure},
    ]{\tabletwo};
    \addplot coordinates {
        (9.4, 6.78)
      };
    \legend {LS, PS, Random};
    \end{axis}
  \end{tikzpicture}\vspace{-10pt}
  \caption{Failure rate (\%) for each measure, comparing linear search (LS) with progression search (PS).}\label{fig:measures}
\end{figure}
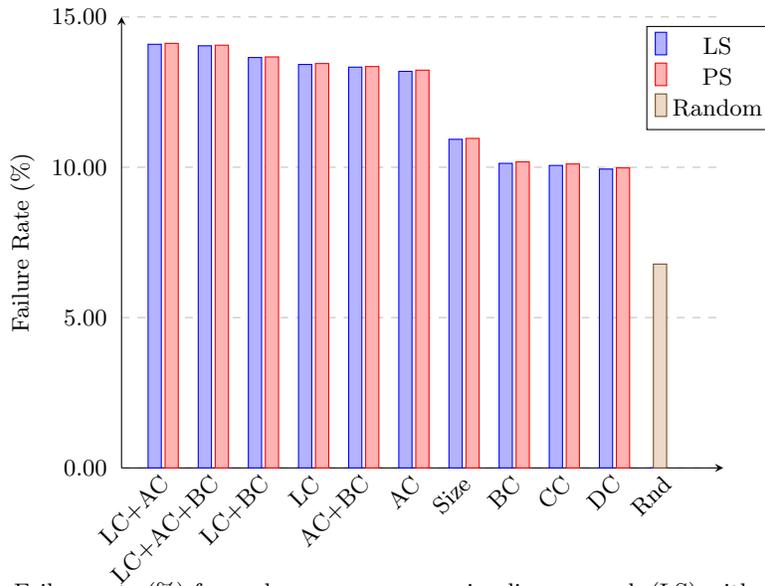

\subsection{RQ3: Linear Combinations of Coverage Measures}

To investigate whether using linear combinations of code coverage measures can further improve the bug finding results, 
we replay the experiment described in \S\ref{subsec:rq1} with the following combinations of coverage measures: 
\etag{1} AC+LC\@; \etag{2} BC+LC\@; \etag{3} AC+BC\@; and \etag{4} AC+BC+LC\@. 

\paragraph{Results.} Figure~\ref{fig:measures} presents the obtained results for the four linear combinations\footnote{LC+AC, LC+BC, AC+BC, and LC+AC+BC} and the five individual code coverage measures presented in Table~\ref{table:results-measures}. For each measure, we give a blue and a red bar, each corresponding to one of the search strategies supported by the \seesaw algorithm. It is easy to observe that the majority of the combinations, i.e., LC+AC, LC+AC+BC, and LC+BC, are able to find more bugs in the students' code than the overall best-performing single measure (LC), with only AC+BC obtaining worse results.

 \section{Related Work}\label{sec:related-work}

\paragraph{Implicit Hitting Sets for Functions Optimization.}
In the last decade, implicit hitting sets (IHS) have been used in many problems with great success.
MaxHS~\cite{DaviesB11,DaviesB13} is a success case of a MaxSAT solver using IHS.
The idea stems from the fact that any minimal correction set is a hitting set of all MUSes.
Once MUSes are enumerated, then a smallest correction set can be obtained by calculating the minimum hitting set of them.
The number of MUSes may be exponential, as such, MaxHS enumerates cores (over approximations of MUSes) and tests whether a correction set is obtained by picking a minimum hitting set of the cores enumerated.
Solving the minimum hitting set problem (MHSP) is as difficult as solving MaxSAT.
However, it has been observed that ILP solvers perform well on MHSP.
Moreno-Centeno and Karp introduce a framework for solving NP-complete problems based on the IHS approach~\cite{karp13}.
Saikko~et~al.\ extend this framework and observe that it is not limited to problems in NP~\cite{matti-kr16}.
In both frameworks, the search considers an \emph{oracle predicate}. The Seesaw algorithm~\cite{seesaw-cp21} generalises the above by considering an \emph{oracle function} rather than just a predicate. Additionally,~\cite{seesaw-cp21} reformulated the precondition of the algorithm by demonstrating that the algorithm is correct as long as the oracle function is anti-monotone.\footnote{Seesaw considers minimization, thus for maximization the oracle function has to be monotone.} 
Monotone predicates
have been studied extensively in the context of SAT~\cite{janota-ai16}.

\paragraph{Test Suit Construction.}
The software engineering community has dedicated a considerable effort to the problem of generating 
effective test suites for complex software systems, exploring topics such as: 
test suite reduction and test case selection~\cite{cruciani:icse:2019,rojas:ssbse:2015,chen:seke:2008,yoo:issta:2007,kitamura:safecomp:2018,jones:transsofteng:2003},  
combinatorial testing~\cite{yamada:icst:2015,yamada:ase:2016,huayao:corr:2019}, 
and a variety of fuzzying strategies~\cite{dart,smart,smash,sage,cute}.
In the following, we focus on the test suite reduction and test case selection problems, which are immediately close
to our own goal, highlighting  
constraint-based approaches. 
Importantly, we are not aware on any works in this field specifically targeted at student projects. 
The testing of such projects has, however, its own specificities when compared to the testing of large-scale industrial software systems. 
In particular, the time constraints on the test generation process are less severe and the code being tested less complex.  

The \emph{test suite reduction problem}~\cite{shi:fse:2015,cruciani:icse:2019,miranda:icse:2018,chen:seke:2008,yoo:issta:2007} is the problem 
of reducing the size of a given test suite while satisfying a given test criterion. 
Typical criteria are the so-called coverage-based criteria, which ensure that the coverage of the reduced test suite is above a
certain minimal threshold. 
The \emph{test case selection problem}~\cite{shi:fse:2015,cruciani:icse:2019,miranda:icse:2018,chen:seke:2008,yoo:issta:2007}  is the dual problem, in that it tries to determine the minimal number of tests to be 
added to a given test suite so that a given test criterion is attained. 
As most of these algorithms are targeted at the industrial setting, they assume severe time constraints on 
the test selection process.
Hence, the vast majority of the proposed approaches for test suite reduction and selection are based on approximate algorithms, 
such as similarity-based algorithms~\cite{cruciani:icse:2019,miranda:icse:2018}, 
which are not guaranteed to find the optimal test suite even when given enough resources. 
In order to achieve a compromise between precision and scalability, the authors of~\cite{chen:seke:2008} proposed 
a combination of standard ILP encodings and  heuristic approaches. 
Finally, the authors of~\cite{kitamura:safecomp:2018} proposed a SAT-based encoding for 
selecting optimal test suites according to the modified condition decision coverage criterion~\cite{Szugyi:2013,jones:transsofteng:2003}. 
They argue that, as this criterion is enforced by safety standards in both the automative and the avionics industries, one is
obliged to resort to exact approaches. 
 \section{Conclusion}\label{sec:conclusion}

We have presented \testselector, a new framework for the automatic selection of optimal test suites for student projects. 
The key innovation of \testselector is that it can be extended with support for new code coverage measures without these 
measures being encoded into the logic of an exact constraint solver. We evaluate \testselector against a benchmark 
comprised of 1,637 real-world student projects, demonstrating that:
\etag{1} it is trivial to extend \testselector with support for new coverage measures and 
\etag{2} the selected test suites outperform randomly selected ones in finding bugs in students' code. 

In the future, we plan to conduct a more thorough investigation on the relation between the characteristics of a project 
and the code coverage measures that are appropriate for it. 
We also plan to integrate \testselector with an existing testing platform, such as Mooshak~\cite{mooshak} or Pandora~\cite{pandora}. 

{\small
\paragraph{Acknowledgements.} 
The authors were supported by Portuguese national funds through Funda\c c\~ao para a Ci\^encia e a Tecnologia 
(UIDB/50021/2020, INESC-ID multi-annual funding program) and projects INFOCOS (PTDC/CCI-COM/32378/2017) 
and DIVINA (CMU/TIC/0053/2021).
The results were also supported by the MEYS within the dedicated program ERC~CZ under the project \emph{POSTMAN} no.~LL1902,
and it is part of the \emph{RICAIP} project that has received funding from the European
Union's Horizon~2020 under grant agreement No~857306.
}
  \bibliographystyle{splncs04}
\bibliography{references}
\newpage
\appendix
\section{Appendix A}\label{sec:appendix-a}

\begin{figure}
  \begin{tabular}{lll}
    \mintinline{c}{BEGIN(char *stmt_t, int id)} & & \begin{tabular}[c]{@{}l@{}}
      Logs the beginning of a control-flow statement with \\
      type \texttt{stmt\_t} and identifier \texttt{id}.
    \end{tabular} \\[10pt]
\mintinline{c}{END(char *stmt_t, int id)} & & \begin{tabular}[c]{@{}l@{}}
      Logs the ending of a control-flow statement with \\
     type \texttt{stmt\_t} and identifier \texttt{id}.
    \end{tabular} \\[10pt]
   \mintinline{c}{BEGIN_FUN(char *name)} & & \begin{tabular}[c]{@{}l@{}}
      Logs the beginning of the function $\mathtt{name}$.
    \end{tabular} \\[5pt]
\mintinline{c}{END_FUN(char *name)} & & \begin{tabular}[c]{@{}l@{}}
      Logs the ending of the function $\mathtt{name}$.
    \end{tabular} \\[5pt]
\mintinline{c}{BLOCK(int id)} & & \begin{tabular}[c]{@{}l@{}}
      Logs the block identifier \texttt{id}.
    \end{tabular} \\[5pt]
\mintinline{c}{GUARD(int expr)} & & \begin{tabular}[c]{@{}l@{}}
      Logs the value \texttt{expr} of a conditional guard.
    \end{tabular} \\[5pt]
\mintinline{c}{COND(int expr)} & & \begin{tabular}[c]{@{}l@{}}
      Logs the value \texttt{expr} of a boolean sub-expression \\ of conditional guard.
    \end{tabular} \\[10pt]
\mintinline{c}{RETURN(char *name)} & & \begin{tabular}[c]{@{}l@{}}
      Logs a \mintinline{c}|return| statement in the function $\mathtt{name}$.
    \end{tabular} \\[5pt]
\mintinline{c}{BREAK(int id)} & & \begin{tabular}[c]{@{}l@{}}
      Logs a \mintinline{c}|break| statement inside a control-flow construct \\ 
      with the identifier $\mathtt{id}$.
    \end{tabular} \\[10pt]
\mintinline{c}{CONT(int id)} & & \begin{tabular}[c]{@{}l@{}}
      Logs a \mintinline{c}|continue| statement inside a control-flow \\ 
      construct with the identifier $\mathtt{id}$.
    \end{tabular} \\[10pt]
\mintinline{c}{DEFAULT(int id)} & & \begin{tabular}[c]{@{}l@{}}
      Traces a \mintinline{c}|default| statement inside a control-flow \\ 
      construct with the identifier $\mathtt{id}$.
    \end{tabular} \\[10pt]
\begin{tabular}[c]{@{}r@{}}
      \mintinline{c}{ARR_DECL(char *name)} \\
    \end{tabular} & & \begin{tabular}[c]{@{}l@{}}
      Logs the static allocation of an array with the static \\ 
      identifier \texttt{name}. 
    \end{tabular} \\[10pt]
    \begin{tabular}[c]{@{}r@{}}
      \mintinline{c}{ARR_WRITE(char *name,} \\
      \mintinline{c}{void *ptr, } \\
      \mintinline{c}{int i)     }
    \end{tabular} & & \begin{tabular}[c]{@{}l@{}}
      Logs an update operation to the index \texttt{i} of the \\ 
      array pointed to by \texttt{ptr} and associated with \\ 
      the static identifier \texttt{name}. 
    \end{tabular} \\[15pt]
\begin{tabular}[c]{@{}r@{}}
      \mintinline{c}{ARR_READ(char *name,} \\
      \mintinline{c}{void *ptr, } \\
      \mintinline{c}{int i)     }
    \end{tabular} & & \begin{tabular}[c]{@{}l@{}}
      Logs a lookup operation of the index \texttt{i} of the \\ 
      array pointed to by \texttt{ptr} and associated with \\ 
      the static identifier \texttt{name}. 
    \end{tabular} \\[15pt]
\begin{tabular}[c]{@{}r@{}}
      \mintinline{c}{STRUCT_REF(char *name, } \\
      \mintinline{c}{void *base, } \\
      \mintinline{c}{void *field)}
    \end{tabular} & & \begin{tabular}[c]{@{}l@{}}
      Logs a pointer dereference operation to the memory \\ 
      region \texttt{base} with the offset \texttt{field}, and associated \\
      with the static identifier \texttt{name}. 
    \end{tabular} \\[15pt]
\mintinline{c}{CALL_MALLOC(unsigned size)} & & \begin{tabular}[c]{@{}l@{}}
    Logs a call to \texttt{malloc} with argument \texttt{size}.
    \end{tabular} \\
  \end{tabular}
  \caption{Complete Coverage Summary API.}
\end{figure}
 
\end{document}